\documentclass[12pt,a4paper]{article}
\usepackage{graphicx}
\usepackage{float}
\usepackage{amsmath,amssymb,amsfonts}
\usepackage{cite}

\newcommand{\bse}{\begin{subequations}}
\newcommand{\ese}{\end{subequations}}
\newcommand{\be}{\begin{equation}}
\newcommand{\ee}{\end{equation}}
\newcommand{\bea}{\begin{eqnarray}}
\newcommand{\eea}{\end{eqnarray}}
\newcommand{\ba}{\begin{array}}
\newcommand{\ea}{\end{array}}

\newcommand{\p}{\partial}

\def\FF{{\mathcal{F}}}

\def\ZZ{{\mathcal{Z}}}

\def\DD{{\mathcal{D}}}
\def\GG{{\mathcal{G}}}
\def\RR{{\mathcal{R}}}

\input amssym.def
\input amssym.tex

\usepackage[colorlinks=true, linkcolor=blue, bookmarks=true, citecolor=red]{hyperref}

\usepackage{geometry}
\begin{document}

\title{Do Horizons Exist?}
\date{}

\maketitle

\hrule
\vspace{5mm}

\begin{center}
\author{Davood Allahbakhshi\footnote{allahbakhshi@ipm.ir} \\\vspace{5mm} School of Particles and Accelerators, Institute for Research in Fundamental Sciences (IPM), P.O.Box 19395-5531, Tehran, Iran}
\end{center}

\vspace{5mm}
\hrule

\linespread{1.5}
\abstract{
\noindent
Gravitational effective action is calculated to second order in transverse momentums for a planar asymptotically anti-de Sitter geometry by gauge fixing method. The first order bulk energy-momentum tensor is calculated. The zeroth order equations of motion are solved and a new black brane-like solution is found. We show that once the contribution of matter quantum modes is taken into account, the horizon of the black brane disappears. This is also correct for BTZ black hole. Our result strengthens the black hole non-existence proposal by Hawking.}

\newpage

\hrule
\tableofcontents
\vspace{5mm}
\hrule

\section{Introduction}

During 1970's, investigating quantum field theories in curved spaces led to some interesting results and hard to solve problems. Works by Fulling \cite{Fulling:1972md},  Davies \cite{Davies:1974th}, Hawking \cite{Hawking:1974sw} and Unruh \cite{Unruh:1976db} showed that time-dependent geometries and black holes produce particles. Also non-inertial observers detect a bath of thermally distributed particles around themselves. On the other hand thermal evaporation of black holes leads to information paradox \cite{Hawking:1976ra}. From Hawking's calculation it seems that a black hole always evaporates to a thermally distributed system of particles, independent of the state of the initial collapsing matter. This makes transition between pure states and mixed states possible which not only destroys the information of the initial system but also is not possible by unitary evolutions of quantum mechanics.

Thermality of Hawking radiation is a result of eternity of the black hole under consideration. One can expect that including the backreaction of the radiation on the black hole will make the radiation non-thermal. These non-thermal deviations may carry information about the initial state of the collapsing matter. But including the backreaction of the radiation on the black hole is not easy. It needs simultaneus calculation of quantum particle production and also solving the equations of motion for geometry.

On the other hand the geometry has some important effects on quantum phenomena. The curved geometry reflects a fraction of the radiation back to the black hole. Also if the space has a boundary, the effects of boundary conditions are very important.

The systematic solution to all these difficulties is using the \emph{gravitational effective action}. This is an effective action for a \emph{general indetermined metric} which can be derived by integrating matter quantum modes out. Solving the equations of motion from such effective action will result classical geometries which include the effects of matter quantum modes. In the mid 1980's an avalanche of work appeared on the subject by many physicists that continues up to now, concerning the effective action, its features and applications.
Some pioneer works and examples of applications are\footnote{The interested reader can find many such articles by following the keywords like \emph{Vilkovisky-DeWitt effective action}. I would like to thank Sergei Odintsov that brought it to my attention.} \cite{Vilkovisky:1984st,Barvinsky:1985an,Ramond:1987td,Kunstatter:1986qd,Rebhan:1986wp,Buchbinder:1992rb,Nojiri:1997sr}.


Since the metric should be general, this action is not achievable in principle! But instead of considering a completely general metric, we can consider the metric as general as possible. A covariant method introduced for expanding the one-loop effective action for asymptotically flat geometries by Barvinsky and Vilkovisky \cite{Barvinsky:1987uw}. They have also calculated the second order terms \cite{Barvinsky:1990up}. Barvinsky, Gusev, Zhytnikov and Vilkovisky also calculated the third order terms in curvature \cite{Barvinsky:1990uq,Barvinsky:1993en}.

But many interesting geometries, including black holes and asymptotically de Sitter and anti-de Sitter spaces, which are being used in different contexts, are not weakly curved and/or asymptotically flat. A very special example is the AdS/CFT correspondence \cite{Maldacena:1997re,Witten:1998qj}. In gauge/gravity setups since the partition function of the bulk with special boundary conditions is the generating functional of the dual field theory \cite{Gubser:1998bc}, having such gravitational effective action for asymptotically AdS spaces, can be very useful. There is another special point about the AdS black holes and black branes. For studying the effects of the radiation on the black hole by using the effective action, we need to calculate the action for a time-dependent geometry which is an extra difficulty. In AdS space, black holes (branes) can be in thermal equilibrium with their radiation, since the boudary of the space is repulsive. It means the final equilibrated system is the solution of a time-independent effective action which is easier to be computed.

Recently a method is proposed for doing quantum calculations in planar geometries, independent of their curvature and asymptotic behaviour, named gauge fixing method \cite{Allahbakhshi:2016lfg}. In this paper we use this method to calculate the gravitational effective action for an asymptotically AdS geometry to second order in transverse momentums. Then we calculate the energy-momentum tensor to first order. Also by solving the equations of motion to zeroth order we find a geometry very similar to the exterior region of an AdS black brane but without the horizon! Disappearance of the horizon, due to quantum modes, in the solution that we found in this work, can be interpreted in line with the \emph{black hole non-existence} proposals by Hawking \cite{Hawking:2014tga}, Frolov \cite{Frolov:2014wja} and others. Also there are some papers which suggest that black holes can not even form \cite{Kawai:2013mda,Vaz:2014era,Abedi:2015yga,Ho:2015vga}. Finally we finish the paper by summary and discussion.

\section{Including Quantum Modes}
As mentioned, for studying the effect of Hawking radiation on the evaporation process of a black hole, we need to study the dynamics of evaporation with all side effects such as backreaction, greybody factor, etc. For this we need to calculate the \emph{gravitational effective action} for a \emph{highly curved time-dependent geometry.} Then we should try to find a time-dependent solution for this effective action; what we are very far from it now. But we can consider an AdS black brane which is equilibrated with its radiation and so we only need to calculate a \emph{time-independent} gravitational effective action.

In this section we try to study the effect of quantum modes of the matter scalar field on the classical background geometry. The quantum modes and the geometry are equilibrated and so we will have a time-independent geometry. For this we use the gauge fixing method which is introduced in \cite{Allahbakhshi:2016lfg}. Finally we will solve the zeroth-order equations of motion and find a \emph{black brane-like} solution \emph{without horizon.} We will see that as soon as we add this zeroth-order correction, the horizon will be disappeared.

\subsection{Gauge Fixing Method}

For including the effects of matter quantum modes we need to calculate the effective action. Consider an asymptotically locally AdS space equipped with a metric $g_{\mu\nu}$ and suppose that the geometry is \emph{near planar}
\be
g_{\mu\nu}(z,x)=\hat{g}_{\mu\nu}(z)+h_{\mu\nu}(z,x).
\ee
So the action is
\be
S = \int{\sqrt{-g(z,x)}\big[g^{zz}\p_z\phi^*\p_z\phi + g^{z\mu}(\p_z\phi^*\p_\mu\phi + \p_\mu\phi^*\p_z\phi) + g^{\mu\nu}\p_\mu\phi^*\p_\nu\phi\big]dz\;d^dx}.
\ee
For a near planar geometry, we can write the first term in the action above in this form
\be
\sqrt{-g(z,x)}g^{zz}\p_z\phi^*\p_z\phi = \sqrt{-\hat{g}(z)}\hat{g}^{zz}\p_z\phi^*\p_z\phi + \delta X(z,x) \p_z\phi^*\p_z\phi .
\ee
The next step is changing the coordinate $z$ to $\xi$ in a way that $\sqrt{-\hat{g}(z)}\hat{g}^{zz} \rightarrow X=\sqrt{-\tilde{g}(\xi)}\;\tilde{g}^{\xi\xi} = 1$. This coordinate transformation can be found by computing the integral below
\be
d\xi = \frac{dz}{\sqrt{-\hat{g}(z)}\;\hat{g}^{zz}} \Rightarrow \xi = \int{\frac{dz}{\sqrt{-\hat{g}(z)}\;\hat{g}^{zz}}}.
\ee
Near boundary analysis of this new coordinate is simple for an AdS space. Near the boundary we have
\be
\sqrt{\hat{g}(z)}\;\hat{g}^{zz}\sim z^{-(d-1)}.
\ee
So the near boundary behaviour of $\xi$ is
\be
\xi = \frac{z^d}{d}\big( 1+z+... \big).
\ee
It means that $\xi > 0$ and the boundary is at $\xi = 0$. The action will be
\be\label{action1}
S = \int{\big[\p_\xi\phi^*\p_\xi\phi +\delta \tilde{X} \p_\xi\phi^*\p_\xi\phi + A^\mu(\p_\xi\phi^*\p_\mu\phi + \p_\mu\phi^*\p_\xi\phi) + B^{\mu\nu}\p_\mu\phi^*\p_\nu\phi\big]d\xi\;d^dx},
\ee
where
\bea
\sqrt{-\tilde{g}}\;\tilde{g}^{\xi\xi} &&= 1\cr\cr
A^\mu = \sqrt{-\tilde{g}}\;\tilde{g}^{\xi\mu} &&= \sqrt{-g}\;g^{z\mu}\cr\cr
B^{\mu\nu} = \sqrt{-\tilde{g}}\;\tilde{g}^{\mu\nu} &&= \sqrt{g\;\hat{g}}\; \hat{g}^{zz}\;g^{\mu\nu}
\eea
and $\delta\tilde{X}$ is the transformed version of $\delta X$. In the action \ref{action1}, the first term will be the \emph{kinetic term} and other terms will be considered as \emph{interaction terms}.
After integrating the first term by parts we have
\be
S = \int{\big[-\phi^*\p^2_\xi\phi +\delta X \p_\xi\phi^*\p_\xi\phi + A^\mu \; (\p_\xi\phi^*\;\p_\mu\phi + \p_\mu\phi^*\p_\xi\phi) + B^{\mu\nu} \; \p_\mu\phi^*\p_\mu\phi\big]d\xi\;d^dx}.
\ee
Here we have dropped the sign $\tilde{•}$ over $\delta X$, just for simplicity. By a wick rotation $t \rightarrow i\tau $ we have the Euclidean action
\be\label{FA1}
S= i S_E = i\; \int{\big[-\phi^*\p^2_\xi\phi +\delta X \p_\xi\phi^*\p_\xi\phi + \hat{A}^\mu \; (\p_\xi\phi^*\;\p_\mu\phi + \p_\mu\phi^*\p_\xi\phi) + \hat{B}^{\mu\nu} \; \p_\mu\phi^*\p_\mu\phi\big]d\xi\;d^d\hat{x}},
\ee
where $\hat{A},\hat{B}$ and $d^d\hat{x}$ are the Euclidean functions and measure. For simplicity we drop this hat sign, since they will be turned back to their original form at the end of the calculation. Now we want to calculate the path integral
\bea
\ZZ &&= \int{e^{iS}\DD\phi}=\int{e^{-S_E}\DD\phi}\cr
&&=\int{e^{\int{\big[\phi^*\p^2_\xi\phi -\delta X \p_\xi\phi^*\p_\xi\phi - A^\mu \; (\p_\xi\phi^*\;\p_\mu\phi + \p_\mu\phi^*\p_\xi\phi) - B^{\mu\nu} \; \p_\mu\phi^*\p_\mu\phi\big]d\xi\;d^dx}}\DD\phi}.
\eea

Before calculating the Green's function and its derivatives, let us have a look at the spectrum. The solution to the first term is
\be
\phi (\xi) = a\xi + b.
\ee
Near the boundary it has the form
\be
\phi (\xi) = a\;z^d\big( 1+z+...\big) + b.
\ee
These two solutions, with coefficients $a$ and $b$, are in fact the \emph{normalizable} and \emph{non-normalizable} modes of a free, massless scalar field in AdS space. On the other hand the spectrum of the first term is 
\be
(- q^2,(2\pi)^{-\frac{d+1}{2}}\;e^{iq\xi+ikx}).
\ee
We like the fileds to be zero at the boundary $\xi = 0$. With this boundary condition the eigen-functions are
\be
(2\pi)^{-\frac{d}{2}}(\pi)^{-\frac{1}{2}}\;sin(q\xi)\;e^{ikx},
\ee
so the one dimensional Green's function is
\be\label{gf1}
\GG (\xi ,x; \xi ' ,y)=-C[\phi^*(\xi ,x),\phi(\xi ' ,y)] = - \frac{\delta^d(x-y)}{\pi}\int{\frac{sin(q\xi)sin(q\xi ')}{q^2}dq},
\ee
where $C[\bullet,\bullet]$ refers to the \emph{contraction} of fields. The explicit form of the Green's function is
\be
\GG (\xi ,x; \xi ' ,y) = \frac{\delta^d(x-y)}{2}\big[ \; |\xi - \xi '| - |\xi + \xi '| \; \big]
\ee
and its derivatives are
\bea
\p_\xi \GG (\xi ,x; \xi ' ,y) &&= \frac{\delta^d(x-y)}{2}\big[\; sign(\xi - \xi ') - sign(\xi + \xi ') \; \big]=-\delta^d(x-y)\theta(\xi ' - \xi)\cr
\p_\xi \p_{\xi '} \GG (\xi ,x; \xi ' ,y) &&= -\delta^d(x-y)\delta(\xi ' - \xi),
\eea
where $\theta$ is the \emph{Heaviside theta} function. And also we have the special case
\be
C[\p_\xi \phi^*(\xi ,x)\;,\; \phi(\xi ,y)]=C[\p_\xi \phi(\xi ,x)\;,\; \phi^*(\xi ,y)] = \frac{\delta^d(x-y)}{2}sign(\xi).
\ee
One interesting point about this Green's function is that
\be
\GG (\xi ,x; \xi ,y) = -\xi \; \delta^d(x-y),
\ee
which is not divergent. Note that in these calculations we have used the positivity of both $\xi$ and $\xi '$. With Green's function and its derivatives in hand we can calculate any quantum mechanical quantity including the \emph{gravitational effective action}, which in the context of AdS/CFT is nothig but the \emph{generating functional} for both the bulk and boundary theories.

\subsection{Second Order Effective Action}

To have a nice day (!) let us move from transverse coordinates to transverse momentums. The action in terms of transverse momentums is
\bea
S = \int &&\big[ \phi ^* \p _\xi^2 \phi - \delta \tilde{X} (\xi , p-k) \p _\xi \phi ^*(\xi , k) \p _\xi \phi (\xi , p) \cr\cr 
&&- \tilde{A}(\xi , p-k)\; \phi(\xi , p)\p _\xi \phi ^*(\xi , k) - \tilde{A}^*(\xi , p-k)\; \phi ^*(\xi , p)\p _\xi \phi(\xi , k)\cr\cr 
&&- \tilde{B}(\xi , p-k) \phi ^*(\xi , p)\phi (\xi , k) \big] \;d\xi d^dk\;d^dp ,
\eea
where
\bea
\delta \tilde{X}(\xi , p-k) &&= \int{\delta X(\xi , x)\;e^{i(p-k).x}\;d^dx}\cr\cr
\tilde{A}(\xi , p - k) &&= \int{ip_\mu A^\mu (\xi , x)\; e^{i(p-k).x}\;d^dx}\cr\cr
\tilde{B}(\xi , p - k) &&= \int{p_\mu k_\nu B^{\mu\nu} (\xi , x)\; e^{i(p-k).x}\;d^dx}.
\eea
Now the contraction of the fields are
\bea
C[\phi^*(\xi , p),\phi(\xi ' , k)]=-\GG (\xi , p ; \xi ' , k)=-\frac{\delta^d(p-k)}{2}\big[ |\xi - \xi '|-|\xi + \xi '| \big]
\eea
and the derivatives of the Green's function with respect to \emph{longitudinal coordinate} ($\xi$), are as before. Now it is easy to calculate the first and second order terms of the \emph{gravitational effective action}. Let us start from $\delta X$ term to all orders. One can show easily that the result is the same as in \cite{Allahbakhshi:2016lfg}
\be\label{X}
\delta(0) \; \int{ln[1+\delta X]d\xi d^dx}.
\ee
Next term is the first order term in $A$ but \emph{all orders} in $\delta X$. One can show that the result is
\be\label{AX}
-\frac{1}{2}\int{\big(\tilde{A}(\xi , p-k)+\tilde{A}^*(\xi , k-p)\big)\FF[F(\delta X(\xi)) , k-p]\;d\xi d^dkd^dp},
\ee
where
\be
F(x)=\frac{1}{1+x}
\ee
and $\FF$ is the Fourier transformation
\be
\FF[f(\xi),p]=\int{f(\xi , x)e^{ip.x}d^dx}.
\ee
You can check that for an exactly planar geometry at the gauge point $X=1$, where $\delta X = 0$, this first order term is zero. Second order terms are in the forms below\footnote{We have dropped the quantum fields $\p_M\phi^* \p_N \phi$ in these expressions, just for simplicity.}
\bea
-&&\int{e^{\phi^*\p_\xi^2 \phi}B \;\DD \phi}\cr\cr
-\sum_{n=1}^{\infty}\frac{(-1)^n}{n!}&&\int{e^{\phi^*\p_\xi^2 \phi}B\delta X^n \;\DD \phi}\cr\cr
\frac{1}{2}&&\int{e^{\phi^*\p_\xi^2 \phi}A^2 \;\DD \phi}\cr\cr
\sum_{n=1}^{\infty}\frac{1}{2}\frac{(-1)^n}{n!}&&\int{e^{\phi^*\p_\xi^2 \phi}A^2\delta X^n \;\DD \phi}\cr\cr
\sum_{N=2}^{\infty}\sum_{n=1}^{N-1}\frac{1}{2}\frac{(-1)^{N}}{(N)!}&&\int{e^{\phi^*\p_\xi^2 \phi}A\delta X^n A\delta X^{N-n} \;\DD \phi}.
\eea
The first term is so simple and its result is
\be\label{B}
-\int{\xi\;\tilde{B}(\xi , 0)\;d\xi}.
\ee
For second term there are $n!$ combinations with the same value and the result is
\be\label{BX}
-\int{\FF[F(\delta X(\xi ')),k-p]\tilde{B}(\xi , p-k) \theta(\xi - \xi ') \;d\xi d\xi ' dpdk},
\ee
where
\be\label{Ffunction}
F(x)=\frac{1}{1+x}-1.
\ee
It is easy to show that the third term is
\be\label{AA}
+\int{\xi\;\tilde{A}(\xi , p-k)\tilde{A}^*(\xi , p-k)\;d\xi dpdk},
\ee
where we have used $\GG(\xi , \xi) = -\xi$.

Also you can show that the fourth term is
\bea\label{AAX}
&&\int{\tilde{A}(\xi , p-k)\tilde{A}^*(\xi , q-k)\FF[F(\delta X(\xi ')),q-p]\theta(\xi - \xi ')\;d\xi d\xi ' dpdkdq}\cr\cr
+&&\int{\xi \; \tilde{A}(\xi , p-k)\tilde{A}^*(\xi , p-q)\FF[F(\delta X(\xi)),k-q]\;d\xi dpdkdq}.
\eea
$F$ is again the function \ref{Ffunction}.

Finally you can verify that the last term is
\bea\label{AXAX}
+\int{\FF[F(\delta X(\xi)) , p-k]\tilde{A}(\xi ', k-w)\FF[F(\delta X(\xi ')), w-t]\tilde{A}^*(\xi ' , p-t) \theta (\xi ' - \xi) \; d\xi d\xi ' dpdkdwdt}\cr\cr
+\int{F(-\delta X(\xi ,x)\delta X(\xi ',y))\;A^\mu(\xi ',x) A^\nu(\xi ',x)\;k_\mu p_\nu \; \theta(\xi ' - \xi) e^{i(p-k).x}e^{-i(p-k).y}d\xi d\xi 'dxdydpdk}.\cr\cr
\eea
Now we can move back to position space. The term \ref{X} is already in position sapce. The term \ref{AX} is
\be\label{1st-order}
-\frac{1}{2}\int{A^\mu(\xi , x)F(\delta X(\xi , y))i(p_\mu -k_\mu) e^{i(p-k).(x-y)} \;d\xi dxdydpdk   },
\ee
where here $F$ is the function \ref{Ffunction}. And second order terms will be
\bea
\ref{B}\rightarrow &&-\int{\xi B^{\mu\nu}(\xi , x) \; p_\mu k_\nu \; d\xi dxdpdk }\cr\cr
\ref{BX}\rightarrow &&-\int{F(\delta X(\xi ,x))B^{\mu\nu}(\xi ', y)\; p_\mu k_\nu \; \theta(\xi ' - \xi ) \; e^{i(p-k).(x-y)}\;d\xi d\xi ' dxdydpdk}\cr\cr
\ref{AA}\rightarrow &&+\int{\xi A^\mu(\xi , x)A^\nu(\xi , x)\;p_\mu p_\nu\;d \xi dx dp}\cr\cr
\ref{AAX}\rightarrow &&+\int{F(\delta X(\xi , x))A^\mu(\xi ',y)A^\nu(\xi ',y)\;p_\mu k_\nu \; \theta(\xi ' - \xi ) \; e^{i(p-k).(x-y)}\;d\xi d\xi ' dxdydpdk }\cr\cr
&&+\int{\xi F(\delta X(\xi , x))A^\mu (\xi , x)A^\nu (\xi , x)\; p_\mu p_\nu \; d\xi dxdp}\cr\cr
\ref{AXAX}\rightarrow &&+\int{F(\delta X(\xi , x))F(\delta X(\xi ', y))A^\mu(\xi ', y)A^\nu(\xi ' , y)\;k_\mu p_\nu  \;\theta (\xi ' - \xi)\; e^{i(p-k).(x-y)} \; d\xi d\xi ' dxdydpdk}\cr\cr
&&+\int{F(-\delta X(\xi ,x)\delta X(\xi ',y))A^\mu(\xi ',y) A^\nu(\xi ',y)\;k_\mu p_\nu \;\theta(\xi ' - \xi)\; e^{i(p-k).(x-y)}d\xi d\xi 'dxdydpdk}.\cr\cr &&\;
\eea
Some of these terms have divergencies in them and need to be renormalized.

\subsection{First Order Energy-Momentum Tensor}
In previous part we calculated the effective action to second order in transverse momentums but all orders in $\delta X$. So we can replace $\delta X$ with $X-1$, where $X$ is not necessarily close to 1, so we have replacements below
\bea
ln(1+\delta X) &&\rightarrow ln(X)\cr\cr
F(\delta X)=\frac{1}{1+\delta X}-1 &&\rightarrow F(X)=\frac{1}{X}-1.
\eea
By using the variations below to first order in $h^{AB}$
\bea
X &&= \sqrt{-g}\;g^{\xi\xi}\approx \sqrt{-\hat{g}}\;\big[\hat{g}^{\xi\xi}+h^{\xi\xi}-\frac{1}{2}\;\hat{g}^{\xi\xi}\;\hat{g}_{AB}\;h^{AB}\big]\cr\cr
A^\mu &&= \sqrt{-g}\;g^{\xi\mu}\approx \sqrt{-\hat{g}}\;\big[ \hat{g}^{\xi\mu}+h^{\xi\mu}-\frac{1}{2}\hat{g}^{\xi\mu}\; \hat{g}_{AB}\;h^{AB} \big]\cr\cr
B^{\mu\nu}&&=\sqrt{-g}\;g^{\mu\nu}\approx \sqrt{-\hat{g}}\;\big[ \hat{g}^{\mu\nu}+h^{\mu\nu}-\frac{1}{2}\hat{g}^{\mu\nu} \; \hat{g}_{AB}\;h^{AB} \big].
\eea
It is easy to see that the variation of the effective action, to first order in transverse momentums or equivalently the first \emph{transverse derivatives} in position space, is
\bea
\delta \Gamma = &&\frac{C_1\;\sqrt{-g}}{X}\big( h^{\xi\xi}-\dfrac{1}{2}g^{\xi\xi}\; g_{AB}\;h^{AB} \big)\cr\cr &&+\dfrac{C_2}{4}\bigg[ \big( A^\mu \; g_{AB}\;h^{AB}-\sqrt{-g} \;[h^{\xi\mu}+h^{\mu\xi}] \big)\;\p_\mu F(X)\cr\cr
&&-\dfrac{2\sqrt{-g}}{X^2}\big( h^{\xi\xi}-\dfrac{1}{2}g^{\xi\xi}\; g_{AB}\;h^{AB} \big)\;\p_\lambda A^\lambda \bigg].
\eea
With this variation, calculating the energy-momentum tensor is trivial
\bea\label{EM-tensor}
\sqrt{-g}\;T_{\xi \xi}&&=\frac{C_1\; \sqrt{-g}}{X}\big(1\;-\frac{1}{2}g^{\xi\xi}g_{\xi\xi} \big)+\frac{C_2}{4}\big[ g_{\xi\xi}A^\lambda\p_\lambda F - \dfrac{2\sqrt{-g}}{X^2} \big( 1-\frac{1}{2}g^{\xi\xi}g_{\xi\xi} \big) \p_\lambda A^\lambda \big]\cr\cr
\sqrt{-g}\;T_{\xi \mu}&&=-\frac{C_1}{2}\;g_{\xi\mu} +\frac{C_2}{4}\big[ -\sqrt{-g}\;\p_\mu F + g_{\xi\mu}\;A^\lambda \p_\lambda F + \frac{1}{X} \;g_{\xi\mu}\; \p_\lambda A^\lambda \big]\cr\cr
\sqrt{-g}\;T_{\mu\nu}&&=-\frac{C_1}{2}\;g_{\mu\nu} +\frac{C_2}{4}\big[g_{\mu\nu}\;A^\lambda \p_\lambda F + \frac{1}{X} \;g_{\mu\nu}\; \p_\lambda A^\lambda \big],
\eea
where $C_1$ and $C_2$ are renormalized couplings. For a $d+1$-dimensional space-time, the trace of the tensor is
\bea\label{Trace}
\sqrt{-g}\;g^{AB}T_{AB}&&=-\frac{(d-1)}{2}\;C_1+\frac{(d-1)C_2}{4}\big[ A^\mu \p_\mu F + \frac{1}{X}\; \p_\mu A^\mu \big]\cr\cr
&&=-\frac{(d-1)}{2}\;C_1+\frac{(d-1)C_2}{4}\;\p_\mu J^\mu ,
\eea
where $J^\mu = A^\mu / X$. For a \emph{planar geometry} the trace is
\be
\sqrt{-g}\;g^{AB}T_{AB}=-\frac{(d-1)}{2}\;C_1.
\ee

Here is good to mention that there are two \emph{planarity preserving} coordinate transformations that we name \emph{transverse} and \emph{longitudinal transformations}. The transverse transformations mix transverse coordinates and just \emph{linear global} transverse transformations are planarity preserving. The longitudinal transformations change the coordinate $\xi$ to another longitudinal coordinate $\rho(\xi)$. The coupling $C_2$ is scalar under both the longitudinal and transverse transformations. The coupling $C_1$ is scalar under transverse transformations but a \emph{covariant vector} under longitudinal transformations.

This can be understood from the energy-momentum tensor but we can show it simply in the trace \ref{Trace}. We can define $J^A = \sqrt{-g}\;g^{\xi A}/X$ where $A$ includes $\xi$ as well as transverse coordinates and so $J^\xi = 1$. Now the trace above can be written in the form below
\be\label{Trace-Invariant}
g^{AB}T_{AB}=-\frac{(d-1)}{2\sqrt{-g}}\;C_1+\frac{(d-1)C_2}{4}\;\bigtriangledown_A j^A,
\ee
where $j^A = g^{\xi A}/X$ and $\bigtriangledown_A$ is the \emph{covariant derivative}. The left hand side is a scalar under all diffeomorphisms so the right hand side should be scalar as well. $\sqrt{-g}$ is a scalar under transverse transformations, for planar geometries, and transforms as a covariant vector under longitudinal transformations. It means that $C_1$  is a scalar under transverse transformations but a covariant vector under longitudinal transformations. The reason is in fact that this coupling is $\lim_{\xi\rightarrow \xi_0}\delta(\xi-\xi_0)$ and so transforms like a covariant vector, since $\delta(\xi - \xi_0)$ transforms in that way. On the other hand $j^A$ is a \emph{contravariant vector} under both transformations which means that $C_2$ is a scalar under them.

The interesting point here is that although just some special transformations are planarity preserving but, the trace \ref{Trace-Invariant} has a complete diffeomorphism invariant form!

\subsection{Semi Black Brane as Zeroth-Order Solution}
In this part we want to solve the equations of motion derived from the zeroth-order effective action. For this it is enough to consider the vacuum expectation value of the energy-momentum tensor to zeroth-order. The equations of motion for an exactly planar geometry are
\bea\label{eff-eqs}
&&\sqrt{-g}\big( \RR_{AB}-(\frac{1}{2}\RR -\Lambda)\;g_{AB} \big)=\sqrt{-g}\;T_{AB}\cr\cr
&&\sqrt{-g}\;T_{\xi \xi}=\frac{C_1}{g^{\xi\xi}}\;-\frac{C_1}{2}g_{\xi\xi}\cr\cr
&&\sqrt{-g}\;T_{\xi \mu}=-\frac{C_1}{2}\;g_{\xi\mu}\cr\cr
&&\sqrt{-g}\;T_{\mu\nu}=-\frac{C_1}{2}\;g_{\mu\nu}.
\eea
Now let us consider $\Lambda <0$ and a $d+1$ dimensional metric of the form
\be\label{my-ans}
ds^2=\frac{1}{\rho(\xi) ^2}\bigg( \frac{f(\xi)}{\rho(\xi) ^{2(d-1)}}\;d\xi ^2-f(\xi)\;dt ^2+d\vec{x}_{d-1}^2 \bigg).
\ee
As is obvious, this metric is at the gauge point $\sqrt{-g}\;g^{\xi\xi}=1$. This is an \emph{AdS black brane-like} metric in $\xi$ coordinate. In fact for black brane we have
\bea
\rho(\xi)=\rho _h \; (1-e^{-d\;\xi})^{1/d}\cr\cr
f(\xi)=1-\frac{\rho(\xi)^d}{\rho _h^d}=e^{-d\;\xi}.
\eea
Just for simplicity, let us consider a 5-dimensional space ($d=4,\;\Lambda=-6$). Plugging the ansatz \ref{my-ans} in equations \ref{eff-eqs} leads to three equations
\bea
\xi\xi &&\rightarrow \; C_1 f \rho^8 +3 \rho^7 f' \rho '-12 f \rho^6 \rho '^2+12 f^2 =0 \cr\cr
tt &&\rightarrow \;  f \rho^6 \left(-C_1 \rho^2 +6 \rho \rho ''+6 \rho '^2\right)-3 \rho^7 f' \rho '+12 f^2 =0 \cr\cr
ii &&\rightarrow \; f^2 \rho^6 \left(C_1 \rho^2 -6 \rho \rho ''-6 \rho '^2\right)- \rho^8 f'^2+ f \rho^7 \left(\rho f''+3 f' \rho '\right)-12 f^3=0.
\eea
In these expressions $^\prime$ refers to derivative with respect to $\xi$. Since the equations above do not have explicit dependence on $\xi$, we can change the variable of the equations from $\xi$ to $\rho$. If we define $P= \big( \rho^\prime \big)^2$, the equations will be
\bea
&& C_1 \rho ^8 f+3 \rho ^7 P \dot{f}-12 \rho ^6 f P+12 f^2 =0 \cr\cr
&& \rho ^6 f \left(-C_1 \rho ^2+3 \rho  \dot{P}+6 P\right)-3 \rho ^7 P \dot{f}+12 f^2 =0 \cr\cr
&& \rho ^6 f^2 \left( 2 C_1 \rho^2 -6 \dot{P}\rho -12 P \right)-2 \rho ^8 P \dot{f}^2+\rho ^7 f \left(\rho  \dot{f} \dot{P}+2 P \left(\rho \ddot{f}+3 \dot{f}\right)\right)-24 f^3 =0,\cr &&\;
\eea
where $\dot{}$ refers to derivative with respect to $\rho$. The first equation is an algebraic equation for $P$ and its solution is
\be
P=\frac{-C_1 \rho ^8 f(\rho )-12 f(\rho )^2}{3 \rho ^6 \left(\rho  \dot{f}(\rho )-4 f(\rho )\right)}.
\ee
After finding $f(\rho)$, we can solve this equation to find $\rho(\xi)$. For $C_1 \neq 0$, it is easy to check analytically that $\xi (\rho)\sim \rho^d$ for $\rho \ll \rho_h$ and $\xi (\rho)\sim ln(\rho)$ for $\rho \gg \rho_h$ and we checked \emph{numerically} that $\xi (\rho)$ is a smooth, differentiable function consistent with the mentioned analytic asymptotic behaviours. Substituting the $P$ above into the remaining equations leads to the same differential equations, which means that these two functions $\rho$ and $f$ are sufficient to solve all equations. The final equation for $f$ that should be solved is
\be\label{f-eq}
\rho  f \left(C_1 \rho ^8+12 f\right) \ddot{f}+\dot{f} \left(-C_1 \rho ^9 \dot{f}+C_1 \rho ^8 f-36 f^2\right)=0.
\ee
The implicit analytic solution to this equation is
\be\label{implicit-sol}
\log (f(\rho ))=-\frac{1}{\sqrt{1+\frac{C_1 \rho_h^8}{3}}}\left(2 \tanh ^{-1}\left(\sqrt{1+\frac{4 \rho_h^8 f(\rho )}{\rho ^8 \left(1+\frac{C_1\rho_h^8}{3}\right)}}\right) + i \pi \right).
\ee
When the coupling $C_1$ is zero, which means the right hand side of the equations are zero (ignoring the quantum modes), this equation can be solved simply and the solutions are
\bea
f&&=1-\frac{\rho^4}{\rho_h^4}\cr\cr
f&&=1+\frac{\rho^4}{\rho_h^4}.
\eea
The first one is the black brane solution and the other one diverges as $\rho$ increases. When the coupling $C_1$ is not zero and so we have the effect of quantum modes, the solution is not simple but, we can still extract some features of the solution from this equation. Later we solve the differential equation numerically and find the solution. One important point here is that we have a one parameter family of solutions labled by $\rho_h$, which at zero coupling is the position of the horizon of the black brane.

\emph{\textbf{- Analytic Analysis}}\\
We found that the positive $f(\rho)$ is only consistent with positive value of the coupling, so let us consider $C_1 \geq 0$. In this case the $\tanh ^{-1}$ has an imaginary part equals to $-i \pi /2$, since its argument is larger than 1. This imaginary value is canceled by $i \pi$ in the equation. Also we need $f=1$ at the boundary of the space $\rho = 0$. When the coupling is zero, the function $f$ can become negative beyond some $\rho$ which is in fact $\rho_h$, since the imaginary part of the \emph{logarithm} can be canceled again by the same $i \pi$ in the equation and $\tanh ^{-1}$ does not have any imaginary part since its argument is less than 1. It means that the function $f$ can be either positive or negative. But when the coupling is not zero, the function $f$ can not change sign and we will see this in numerical solution.

On the other hand it is easy to see that the near boundary behaviour of $f$, for non-zero coupling, is
\be
f \simeq 1 \pm \frac{\rho ^4}{\rho _h^4}+...
\ee
We will use this behaviour to solve the equation numerically.

\begin{figure}
\centering
\includegraphics[scale=.4]{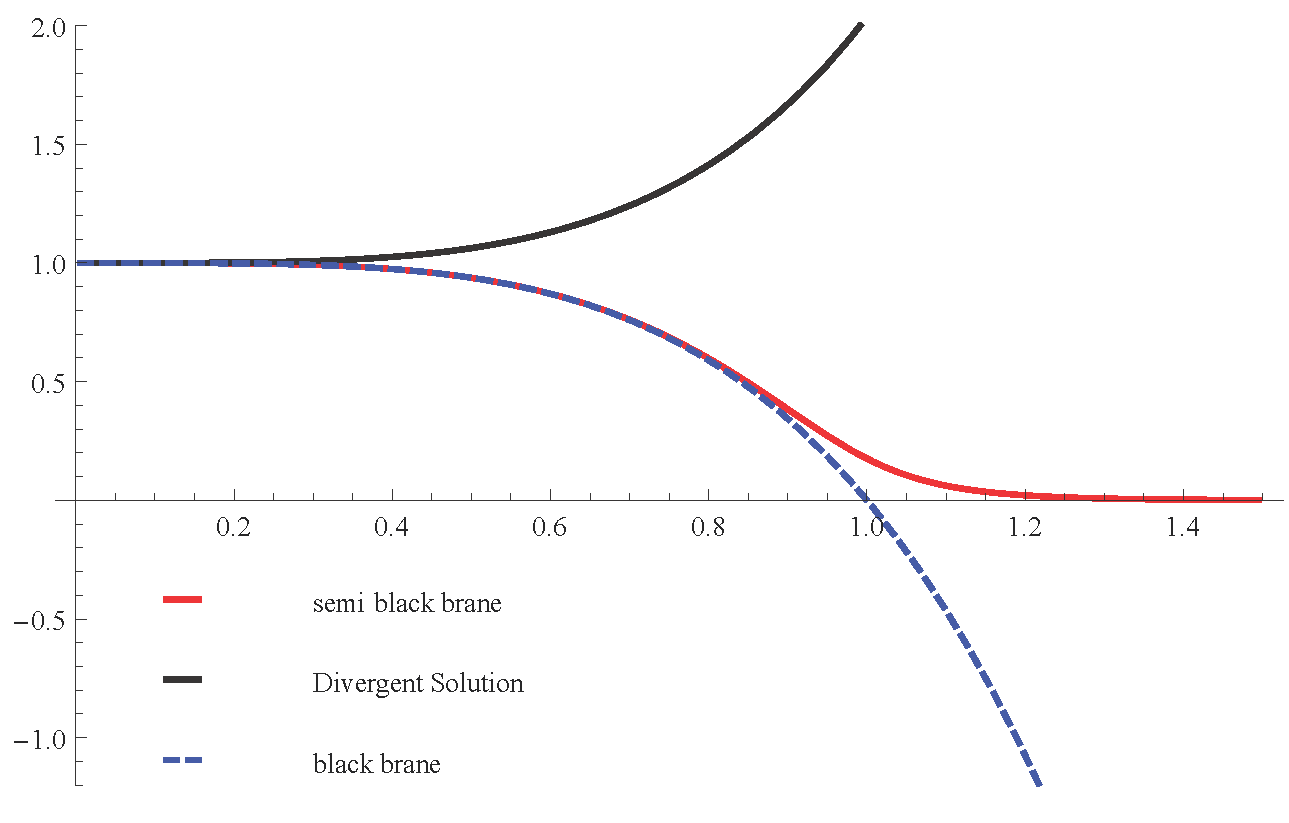}
\includegraphics[scale=.5]{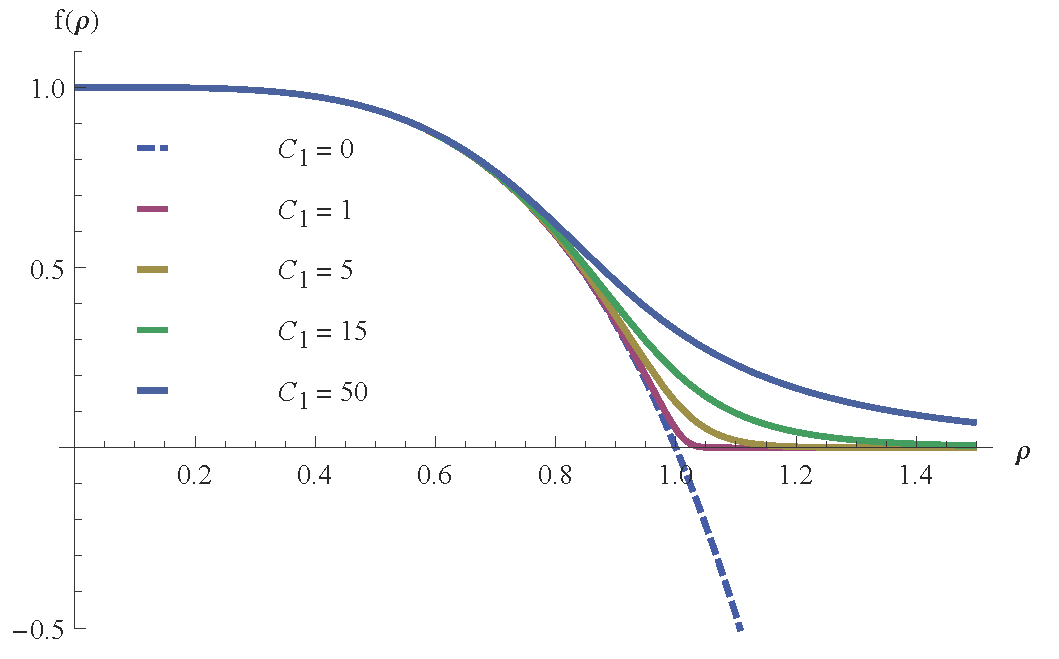}
\caption{\label{pic1}(left) Plot of $f(\rho)$. (right) Plot of $f(\rho)$ of black brane (dashed blue) and semi black brane for different values of $C_1$ with $\rho_h=1$.}
\end{figure}

\emph{\textbf{- Numerical Solution}}\\
Solving the equation \ref{f-eq} numerically with the mentioned boundary behaviour leads to the fig. \ref{pic1}. There is again a divergent solution and the other solution (the red line) is not a black brane, since $f$ never becomes zero at finite $\rho$ but at $\rho \rightarrow \infty$. In fig. \ref{pic1}, we have also plotted the function $f(\rho)$ for different values of the coupling.

There are some interesting facts in this plot. When the coupling is zero, we have simple black brane solution. The geometry has a coordinate singularity at finite $\rho$, where $f$ becomes zero, which is the horizon and then $f$ becomes negative. Once the coupling is turned on, the negative part of the solution (horizon and the interior region of the black brane) disappears. It can be understood from the equation \ref{f-eq}. If $f(\rho)$ grows slower than $\rho^8$ as $\rho$ increases, then for very large values of $\rho$, the equation becomes
\be
C_1 \left(\rho f\ddot{f} -\rho \dot{f}^2+ f\dot{f}\right)=0.
\ee
Independent of the value of the coupling the general solution is
\be
f(\rho)=\alpha \; \rho^\beta .
\ee
From solution \ref{implicit-sol}, it is easy to see that for $\rho \gg \rho_h$ the function $f(\rho)$ is
\be
f(\rho) = \left( \frac{\rho_h^8}{1+C_1 \rho_h^8 / 3} \right)^{\frac{1}{\sqrt{1+C_1 \rho_h^8 / 3}\;-\;1}} \; \rho^{-\frac{8}{\sqrt{1+C_1 \rho_h^8 / 3}\;-\;1}}
\ee

The other important and interesting point is that for small couplings the geometry is very similar to the exterior region of the black brane. It means that for these values of coupling, the static observer at infinity can not understand the difference between black brane and this solution, by classical probes. beyond $\rho=\rho_h$ the function $f$ is very close to zero and so there is, although not completely black, but a very dark region! From the point of view of the static observer at infinity, an infalling matter when reaches $\rho_h$ will become, although not exactly, but approximately frozen. For these reasons we name it a \emph{semi black brane}. Since there is not any horizon, there is not any \emph{Hawking radiation}, so the parameter $\rho_h$ can not be related to any temperature but it should be related to the total energy of the semi black brane.

\begin{figure}
\centering
\includegraphics[scale=.5]{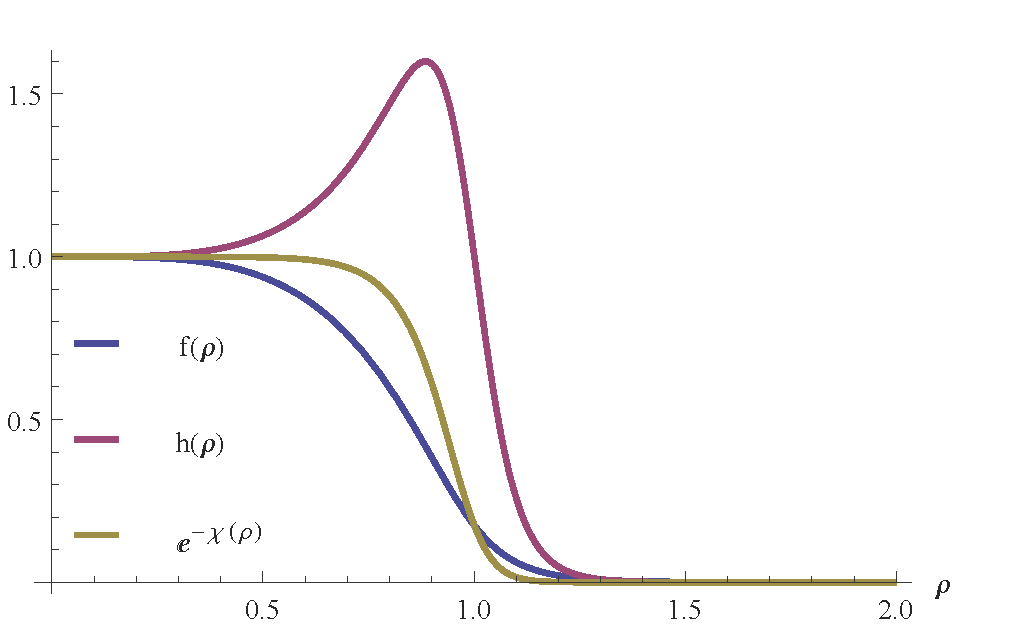}
\includegraphics[scale=.5]{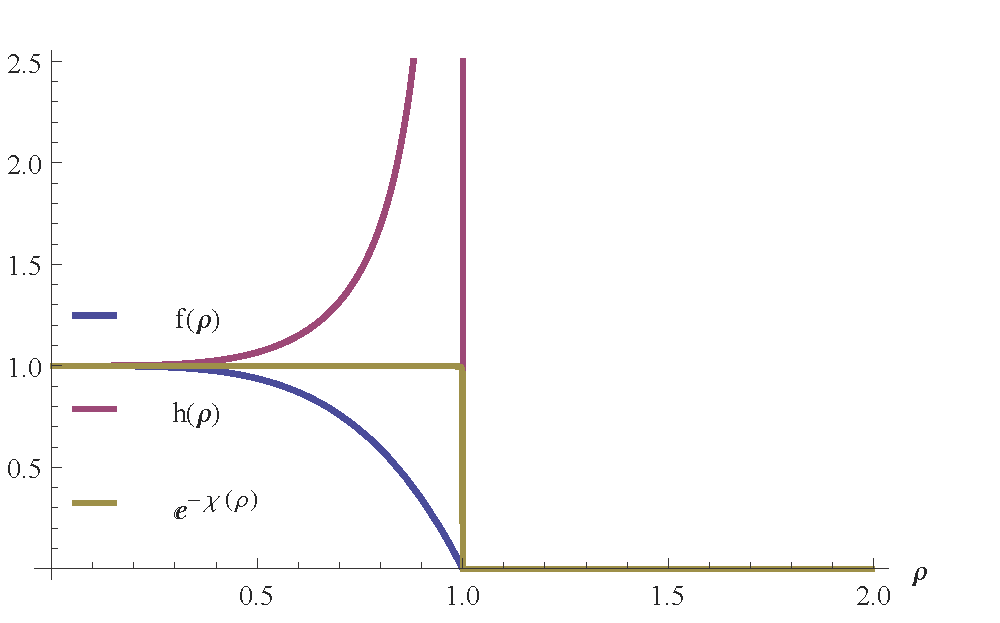}
\caption{\label{pic2}Plot of metric functions for finite (left) and very small (right) value of the coupling.}
\end{figure}

\subsection{Structure of the Semi Black Brane Geometry}
The semi black brane metric in $\rho$ coordinate\footnote{We can use the $\rho$ coordinate with the metric ansatz above, from the begining, to find the solution. But since this ansatz is out of the gauge point $X=1$, we should use $(\p \xi /\p \rho)C_1 = C_1/\sqrt{P}$ instead of $C_1$, as mentioned previously. You can verify that, in $\rho$ coordinate, just by this substitution we derive the differential equations that we found in $\xi$ coordinate.} is
\be\label{my-ans-rho}
ds^2=\frac{1}{\rho ^2}\bigg(h(\rho)\;d\rho ^2-f(\rho)\;dt ^2+d\vec{x}_{d-1}^2 \bigg),
\ee
where
\be
h(\rho) = \frac{f(\rho)}{P(\rho)\rho^{2(d-1)}}.
\ee
We can also write the metric above in this form
\be\label{my-ans-rho}
ds^2 = \frac{1}{\rho ^2}\bigg(\frac{e^{-\chi (\rho)}}{f(\rho)}\;d\rho ^2-f(\rho)\;dt ^2+d\vec{x}_{d-1}^2 \bigg),
\ee
which is similar to a \emph{hairy black brane.} For finite values of the coupling $C_1$, the function $h(\rho)$ is not $1/f(\rho)$ but, it has a maximum value at some $\rho<\rho_h$. At the limit $C_1\rightarrow 0$, this maximum value goes to infinity at $\rho_h$ and the function $h(\rho)$ tends to $1/f(\rho)$ or equivalently $\chi(\rho)\rightarrow 0$.  At this limit the function $f(\rho)$ is 
\bea
\lim _{C_1\rightarrow 0} f(\rho) = 1-\frac{\rho ^4}{\rho _h ^4} \qquad \rho < \rho _h,
\eea
which is that of the exterior region of the black brane. Beyond $\rho_h$ all functions are zero. In fig. \ref{pic2} we have plotted the metric functions for a finite valued and also a very small coupling.

\begin{figure}
\centering
\includegraphics[scale=0.5]{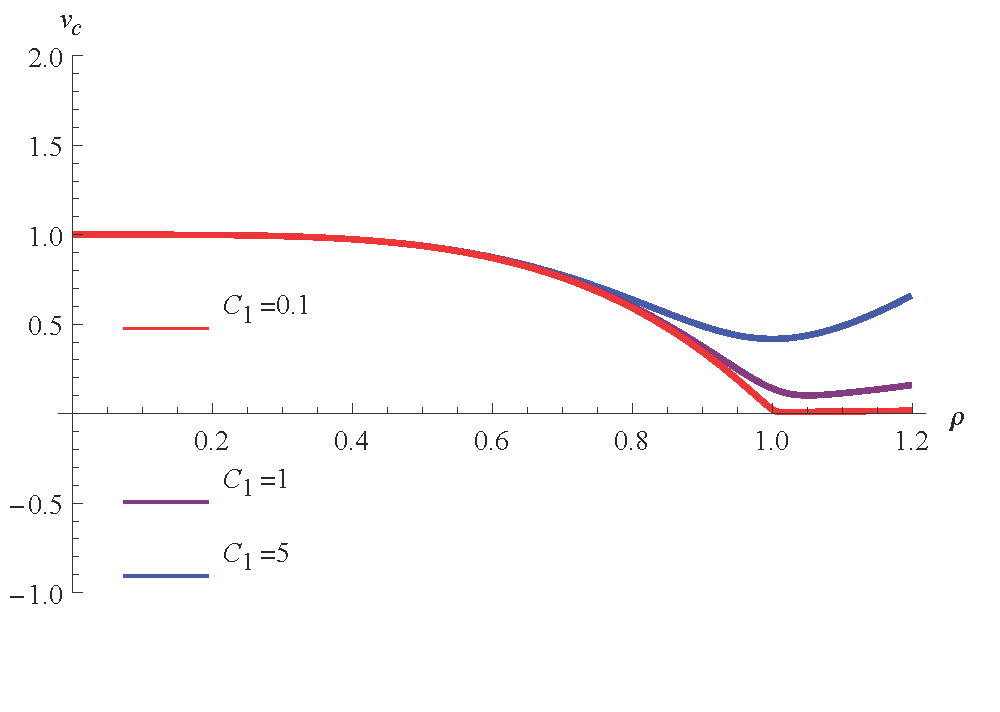}
\includegraphics[scale=0.7]{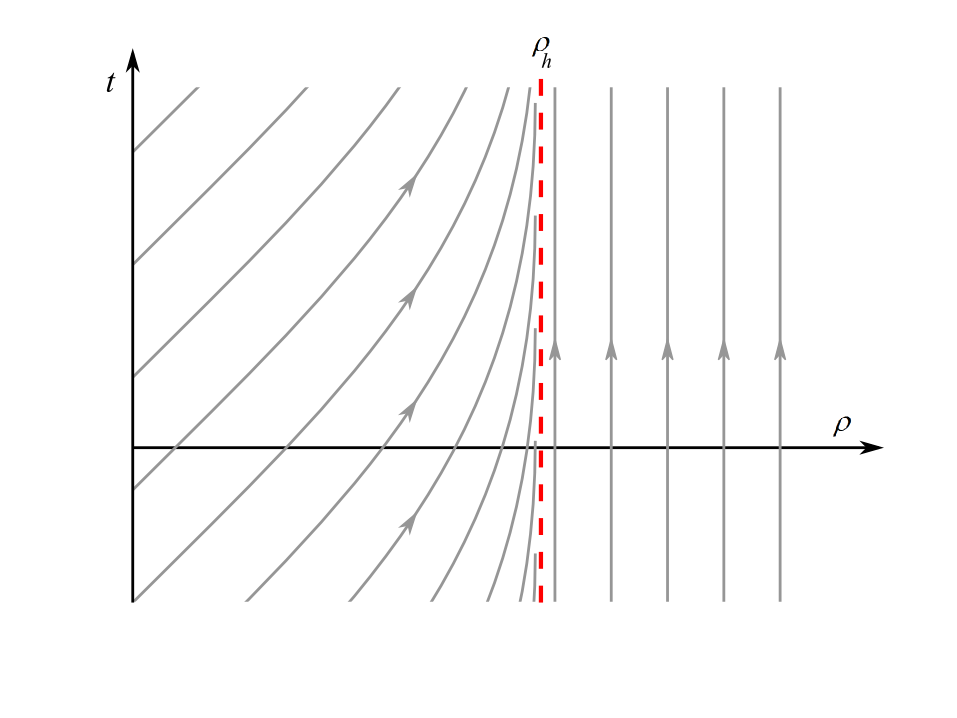}
\caption{\label{pic3}(left) Plot of the coordinate speed of light for different values of the coupling. (right) The space-time plot of light rays, in $\rho$ coordinate, for semi black brane geometry at the limit $C_1 \rightarrow 0$.}
\end{figure}

Since both functions $h$ and $f$ go to zero as $\rho$ goes to infinity, the causal structure of the geometry can not be realized from these plots. For this we need to study other objects, e.g. light rays. The coordinate speed of light in this coordinate is
\be
v_c(\rho) = \rho^{d-1}\sqrt{P(\rho)}.
\ee
This speed is plotted in fig. \ref{pic3} for different values of the coupling $C_1$. For a finite valued coupling the speed starts from $1$ at the boundary and then decreases. It has a minimum value around $\rho_h$ and then increases to infinity as $\rho$ increases.

When $C_1\rightarrow 0$, the geometry becomes the same as the exterior region of the black brane, below $\rho_h$, and so the speed of light tends to $v_c \rightarrow f(\rho)$, similar to the black brane. Beyond $\rho_h$, it becomes \emph{zero} which means that every thing is frozen there from the viewpoint of the observer. This is completely different from \emph{interior region} of the black brane.

\subsection{Semi BTZ black hole}
So far we have shown that the contribution of the leading term in the effective action removes the horizon of \emph{AdS black brane} in all space-time dimensions. But what about black holes? Unfortunately the metric of a black hole is non-planar in general, since the azimuthal components of the metric depend on angles. The only exceptions are three dimensional black holes.

You can check that the zeroth order effective action, that we introduced above, does not have the black hole solution in three dimensions, when the cosmological constant is zero. But when it is not zero there is a \emph{BTZ-like} solution. The metric of a BTZ black hole without rotation and charge is
\be\label{BTZ-form}
ds^2=\frac{1}{z^2}\left[ -f(z)\;dt^2+\frac{dz^2}{f(z)}+d\theta ^2 \right]
\ee
with
\be
f(z)=1-\frac{z^2}{z_h^2},
\ee
which is the solution of the Einstein-Hilbert action.

For finding a BTZ-like solution for the effective action, we consider again a metric of the form
\be
ds^2=\frac{1}{z(\xi)^2}\left[ -f(\xi)\;dt^2+\frac{f(\xi)}{z(\xi)^2}\;d\xi ^2+d\theta ^2 \right],
\ee
which is at the gauge point $X=1$. Equations of motion from the zeroth-order effective action are
\bea
&&C_1 \; f(\xi ) z(\xi )^4+z(\xi )^3 f'(\xi ) z'(\xi )+4 f(\xi )^2-2 f(\xi ) z(\xi )^2 z'(\xi )^2 = 0\cr\cr
&&f(\xi ) z(\xi )^3 \left(2 z''(\xi )-C_1\; z(\xi )\right)-z(\xi )^3 f'(\xi ) z'(\xi )+4 f(\xi )^2 = 0\cr\cr
&&f(\xi )^2 \left(C_1-\frac{2 z''(\xi )}{z(\xi )}\right)-f'(\xi )^2+f(\xi ) \left(f''(\xi )+\frac{f'(\xi ) z'(\xi )}{z(\xi )}\right)-\frac{4 f(\xi )^3}{z(\xi )^4} = 0.
\eea
By changing the independent variable from $\xi$ to $z$, and defining $z'(\xi)^2 = P(z)$, we have
\bea
&& C_1\; z^4 f(z)+z^3 P(z) f'(z)-2 z^2 f(z) P(z)+4 f(z)^2 = 0\cr\cr
&& z^3 f(z) \left(P'(z)-C_1\; z\right)-z^3 f'(z) P(z) +4 f(z)^2 = 0\cr\cr
&& f(z)^2 \left(C_1-\frac{P'(z)}{z}\right)-P(z) f'(z)^2-\frac{4 f(z)^3}{z^4}\cr\cr
&&+f(z) \left(P(z) f''(z)+\frac{1}{2} f'(z) P'(z)+\frac{P(z) f'(z)}{z}\right) = 0.
\eea
The first equation leads to
\be
P(z)=\frac{-C_1\; z^4 f(z)-4 f(z)^2}{z^2 \left(z f'(z)-2 f(z)\right)}.
\ee
Substituting to other equations leads to the equation
\be
z f(z) \left(C_1\; z^4+4 f(z)\right) f''(z)+f'(z) \left(-C_1\; z^5 f'(z)+C_1\; z^4 f(z)-4 f(z)^2\right) = 0,
\ee
with the implicit solution
\be
\log (f(z)) = -\frac{1}{\sqrt{1+C_1 z_h^4}} \left(2 \tanh ^{-1}\left(\sqrt{1+\frac{4 f(z) z_h^4}{(1+C_1 z_h^4)z^4}}\right)+i \pi \right),
\ee
with the same features as before. So both three dimensional AdS-black brane and AdS-black hole are replaced by semi-black brane and semi-black hole.

\section{Summary and Discussion}
In this paper we calculated the gravitational effective action for an asymptotically AdS geometry with the boundary condition $\phi_{boundary}=0$, to second order in transverse momentums. The first order energy-momentum tensor is calculated. We also solved the zeroth order equations of motion and we found that the contribution of the leading order term removes the horizon of the black brane in all space-time dimensions. This is also correct for Three dimensional AdS black brane and BTZ black hole.

\noindent
Here we would like to mention to some points

\textbf{Covariancy} - The effective action should be diffeomorphism invariant since both the \emph{bare action} and the \emph{measure of the path integral} are diffeomorphism scalars. It means that the full effective action should have a covariant form. The gauge fixing method in its perturbative form results a non-covariant effective action. Including higher order terms may restore the covariancy although our truncated effective action is still acceptable as the one in planar coordinates.

\textbf{Holographic Hydrodynamics} - As mentioned, the perturbative gauge fixing method that we used in this paper is a perturbative calculation in transverse momentums. So it can be considered as a hydrodynamical expansion in transverse planes. In an AdS/CFT setup, this will be the hydrodynamical expansion for dual theory. It will be interesting to use this method in that context.

\textbf{Coupling $C_1$} - The contribution of the leading term in effective action is controled by the coupling $C_1$. It is important to remember that the dimension of the coupling is that of the \emph{spatial energy density} ($J/l^d$). We can make it dimensionless by using the dimensionful quantities in the theory which are Newton's and Planck's constants and the speed of light
\be
C_1=\frac{\hbar c}{l_P^{d+1}}\;\hat{C}_1=\frac{c^4}{G_Nl_P^2}\;\hat{C}_1,
\ee
where $l_P$ is the Planck length and $\hat{C}_1$ is a dimensionless coupling. The coupling $C_1$ is related to an \emph{energy density} in the space, comming from \emph{quantum modes} to \emph{zeroth order in transverse momentums}. Disappearence of the horizon is not just the effect of this energy density but the way it couples to the geometry through $lnX$. This special coupling also leads to the trace anomaly \ref{Trace}.

The other point is that the coupling $C_1$, is a constant number, just in gauge point $X=1$. In other coordinates that the metric is out of the gauge point, it is a function which can be found from its transformation. Suppose that the metric is planar in some coordinate $\xi$ and $\rho(\xi)$ is another coordinate. The $C_1$-function in $\rho$ coordinate will be
\be
\tilde{C}_1(\rho) = C_1 \frac{d \xi (\rho)}{d \rho},
\ee
which is a \emph{local energy density} in the new coordinate.

\textbf{The Horizon} - The contribution of the leading term of the effective action removed the horizon of the black brane. Recently it is shown that including the effect of quantum modes through trace anomaly prevents the horizon to be formed \cite{Abedi:2015yga}. This result is comparable to ours. On the other hand, in our calculation, it seems unlikely that the higher order terms of the effective action will restore the horizon. They will modify the effect but should not wash it away completely since they include transverse momentums and can not cancel the term $lnX$.

As we saw in this work, the limit $C_1 \rightarrow 0$ leads to \emph{semi black brane} while $C_1 = 0$ results the \emph{black brane}; the limit is \emph{not Continuous}. It means that if general relativity is a limit of a quantum effective theory, then the picture that it presents from interior region of the black brane is probably not correct. Specially the exterior region of semi black brane, for small couplings, is that of the black brane and so not distinguishable from outside. Our results arise a serious doubt about the existence of horizons when quantum effects of the matter fields and probably gravitons are taken into account. One may also expect this to be the case for asympotically flat black holes. The expectation which is also supported by some previous works \cite{Abedi:2015yga}. Our result may be considered as another supportive observation for black hole non-existence proposal \cite{Hawking:2014tga,Frolov:2014wja}, as mentioned previously.

Of course the results of the current paper should be verified by more investigations, since we have used the leading term of a perturbative non-covariant effective action. Specially, our result may be gauge or method dependent.

\vspace{.5cm}
\emph{Acknowledgement} - I would like to thank A. Davody, M. Safari, N. Abbassi and A. Naseh for usefull discussions.

\vspace{.5cm}
\hrule


\end{document}